\documentclass[a4paper,10pt,twocolumn]{article}
\usepackage[utf8]{inputenc}
\usepackage{multicol,graphicx}
\usepackage{mathptmx}
\usepackage{bm} 
\usepackage[T1]{fontenc}
\usepackage{textcomp}
\usepackage{url}
\usepackage[hidelinks]{hyperref}
\usepackage[T1]{fontenc}
\usepackage[english]{babel}
\usepackage{booktabs}
\usepackage{subfig}

\usepackage[backend=bibtex,style=ieee,doi=false,isbn=false,url=true,maxnames=6,citestyle=numeric-comp,giveninits=true]{biblatex}

\fontfamily{ptm}\selectfont

\bibliography{references.bib}

\usepackage[font=small]{caption}
\usepackage{amsmath, amssymb}

\newcommand{\degree}{^{\circ}}

\newcommand{\mysection}[1]{\vspace{0.4cm} \uppercase{#1} \vspace{0.4cm}}
\newcommand{\mysubsection}[1]{\hspace{10pt}\textit{#1:}}

\begin{document}
	
\setlength{\textfloatsep}{10pt plus 1.0pt minus 2.0pt}	
\setlength{\columnsep}{1cm}

%----------------------------------------------------------------------------------------
% Header (Title, Authors, Facilities)
%----------------------------------------------------------------------------------------

\twocolumn[%
\begin{@twocolumnfalse}
\begin{center}
	{\fontsize{14}{18}\selectfont
	% Title
        \textbf{\uppercase{Towards gaze-independent c-VEP BCI: a pilot study}}\\}
    \begin{large}
        \vspace{0.6cm}
        % Authors
        S. Narayanan\textsuperscript{1}, S. Ahmadi\textsuperscript{1}, P. Desain\textsuperscript{1}, J. Thielen\textsuperscript{1}\\
        \vspace{0.6cm}
        % Facilities
        \textsuperscript{1}Donders Institute, Radboud University, Nijmegen, the Netherlands\\
        \vspace{0.5cm}
        E-mail: \href{mailto:jordy.thielen@donders.ru.nl}{jordy.thielen@donders.ru.nl}
        \vspace{0.4cm}
    \end{large}
\end{center}	
\end{@twocolumnfalse}%
]%

%----------------------------------------------------------------------------------------
% Abstract (160/160 words, no indent)
%----------------------------------------------------------------------------------------

ABSTRACT: 
A limitation of brain-computer interface (BCI) spellers is that they require the user to be able to move the eyes to fixate on targets. This poses an issue for users who cannot voluntarily control their eye movements, for instance, people living with late-stage amyotrophic lateral sclerosis (ALS). This pilot study makes the first step towards a gaze-independent speller based on the code-modulated visual evoked potential (c-VEP). Participants were presented with two bi-laterally located stimuli, one of which was flashing, and were tasked to attend to one of these stimuli either by directly looking at the stimuli (overt condition) or by using spatial attention, eliminating the need for eye movement (covert condition). The attended stimuli were decoded from electroencephalography (EEG) and classification accuracies of $88$\,\% and $100$\,\% were obtained for the covert and overt conditions, respectively. These fundamental insights show the promising feasibility of utilizing the c-VEP protocol for gaze-independent BCIs that use covert spatial attention when both stimuli flash simultaneously.

%----------------------------------------------------------------------------------------
% Introduction
%----------------------------------------------------------------------------------------

\mysection{introduction}

% BCI and EEG
A brain-computer interface (BCI) records its users' brain activity and translates it into a computer command, opening a novel non-muscular channel for communication and control~\cite{wolpaw2002}. Typically, a BCI records brain activity with electroencephalography (EEG) because it is affordable, practical, and non-invasive. 

% c-VEP BCI and speller BCI--- compare performances for SSVEP, P300 
One of the fastest BCIs for communication uses the code-modulated visual evoked potential (c-VEP) as measured in the EEG~\cite{martnezcagigal2021}. The c-VEP is observed during visual stimulation of the user with a pseudo-random sequence of flashes. As each of the presented symbols concurrently flickers with a random but unique sequence of flashes, specific brain activity is evoked when the user attends to one of the symbols. Subsequently, machine learning algorithms infer the attended symbol from the users’ evoked brain activity. Such a visual BCI speller allows its user to select symbols or commands and as such communicate, bypassing most of the motor system~\cite{verbaarschot2021}.

% The challenge of eye movements and gaze-dependence
Unfortunately, an important limitation of a standard visual BCI speller is the requirement of the users' eyes to shift their gaze towards (i.e., fixate on) a target symbol. Because BCI control is fully dependent on eye movements, this poses a major challenge and quickly renders the BCI uncontrollable for people who have lost voluntary control of their eye movements, i.e., people living with late-stage amyotrophic lateral sclerosis (ALS).

% Briefly present related literature on gaze-independent BCI
In the visual domain, several studies have attempted to develop a gaze-independent BCI.
For instance, Blankertz and colleagues developed a BCI speller called the `Hex-o-Spell' that used motor imagery (imagined right hand and right foot movement, i.e., $N=2$ classes) of the user to aid the selection of characters from six hexagonal fields~\cite{blankertz2006}. They reported a typing speed of 2.3--5\,char/min and 4.6--7.6\,char/min, for their two participants respectively.
Interestingly, Treder and Blankertz showed that visual covert spatial attention can also  be used to operate the `Hex-o-Spell' and the `Matrix' speller using the P300 event-related potential (ERP)~\cite{treder2010}. This covert `Hex-o-Spell' outperformed the covert `Matrix' speller, with a classification accuracy of $60$\,\% ($N=36$ classes) and $40$\,\% ($N=30$ classes), respectively. 

Furthermore, work by Treder and colleagues compared the P300-based `Hex-o-Spell', the `Cake Speller', which is similar to the former, and a `Center Speller', where unique geometric shapes with different colors were closely surrounded by characters, and presented centrally on the screen in a sequential fashion~\cite{treder2011a}. A classification accuracy of $91.3$\,\%, $88.2$\,\%, and $97.1$\,\% was reported for the three spellers, respectively ($N=30$ classes). 
Similarly, Chen and colleagues~\cite{Chen2016} used an extension of the P300 oddball paradigm, namely, rapid serial visual presentation (RSVP). The authors used two versions: a colored circles paradigm (CCP), and a dummy face paradigm (DFP). The average performances obtained from the CCP and DFP paradigms were in the range $51.6$\,\% and $73.5$\,\%, respectively.

Additionally, Treder and colleagues, in another instance, focused on using changes in alpha band activity induced by covert attention shifts to classify the direction in which attentional shifts occurred~\cite{treder2011b}. The authors showed that a classification accuracy of $73.65$\,\% was obtained ($N=2$ classes). These results indicate the potential of using alpha activity as a feature for spatial attention decoding in gaze-independent BCIs.

Furthermore, Kelly and colleagues designed a gaze-independent BCI for communication by combining features from the steady-state visual evoked potential (SSVEP) and alpha band modulations to decode covert spatial attention~\cite{kelly2005}. The authors reported an average performance of $70.3$\,\%, $72.8$\,\% and $79.5$\,\% when using the SSVEP, alpha band, or both features in their analysis pipeline, respectively ($N=2$ classes).
% Work by Allison and colleagues~\cite{allison2008} further demonstrated the utility of SSVEPs in designing such gaze-independent BCIs. 
Similarly, Egan and colleagues~\cite{egan2017} aimed for a hybrid gaze-independent speller using the P300 ERP and alpha in addition to the SSVEP. Importantly, adding the P300 response and alpha as additional features in their classification pipeline improved the performance by $17$\,\% to an overall $79$\,\% when compared to the performance using only the SSVEP, achieving $62$\,\% ($N=2$ classes).

% Aims
In this pilot study, we work towards a gaze-independent BCI. The gaze-dependent c-VEP has recently demonstrated exceptional performance, surpassing other evoked paradigms like ERP and SSVEP~\cite{shi2024}. Another study revealed the reliable decoding of c-VEP from peripheral stimulation (away from fixation) compared to direct foveal stimulation (at fixation)~\cite{waytowich2015}.

Our objective is to acquire fundamental insights on the feasibility of decoding the c-VEP in a fully gaze-independent manner. Specifically, participants will use covert spatial attention to concentrate on stimuli, eliminating the need for direct eye movements to foveate on them. In this pilot work, the stimuli were presented sequentially, to assess whether the c-VEP can be decoded from the far periphery, before testing the more complex parallel stimulation case, where stimuli would be presented simultaneously. If successful, this study provides the first steps to a gaze-independent c-VEP BCI, potentially providing a high-speed neuro-technological assistive device for individuals who may not have reliable control of their eye movements.

%----------------------------------------------------------------------------------------
% Materials and Methods
%----------------------------------------------------------------------------------------

\mysection{materials and methods}

\mysubsection{Participants}
Five participants (all male, mean age 31\,years, range 24-50\,years) were included in this study after obtaining written informed consent. Two participants were authors of this study. A pre-screening procedure excluded any participants with a history of epilepsy or brain injury. All participants had normal or corrected-to-normal vision and reported no central nervous system abnormalities. This study was approved by the Ethical Committee of the Faculty of Social Sciences at the Radboud University Nijmegen.

\mysubsection{Materials}
% EEG data
EEG data from 64 Ag/AgCl active electrodes placed according to the international 10-10 system were recorded at 512\,Hz amplified by a Biosemi ActiveTwo amplifier. The data were preprocessed using a notch filter at 50\,Hz and a bandpass filter with a lower cutoff at 1\,Hz and a higher cutoff at 40\,Hz. Subsequently, the data were sliced to trials starting at 500\,ms before stimulus onset until 20\,s after stimulus onset. Finally, the data were downsampled to 120\,Hz, and the 500\,ms pre-stimulus that may have captured filter artefacts due to initial slicing and subsequently filtering were removed.

\begin{figure*}
    \centering
    \begin{subfloat}[]{
        \includegraphics[width=\columnwidth]{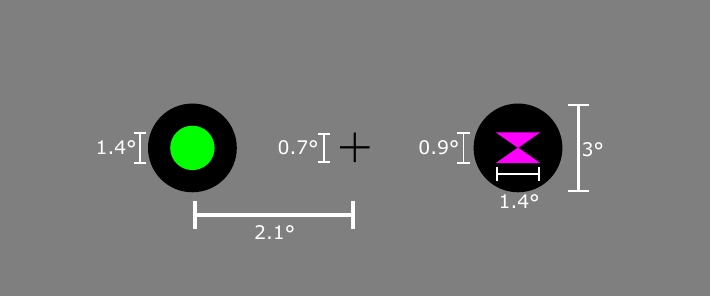}
    }
    \end{subfloat}
    \begin{subfloat}[]{
        \includegraphics[width=\columnwidth]{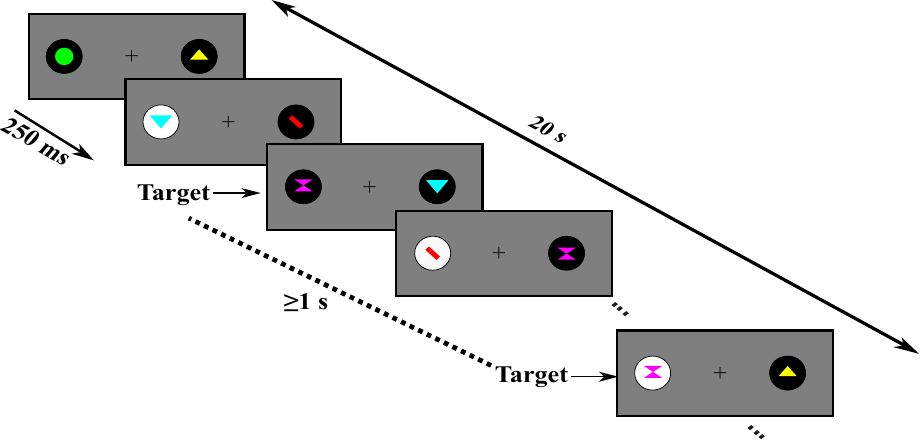}
    }
    \end{subfloat}
     \caption{\label{fig:stimuli}\textbf{Stimulus protocol.} In (a), a graphical representation of the stimulus interface is depicted, featuring two stimuli positioned at $2.1\degree$ on either side of a fixation cross. The stimuli took the form of circles measuring $3\degree$ in both height and width. The fixation cross was $0.7\degree$ for each side. The shapes presented were bound to a maximum height and width of $1.4\degree$ each. The shapes' heights and widths were as follows: green circle ($1.4\degree$ diameter), inverted cyan triangle and yellow triangle ($0.9\degree \times 1.4\degree$), magenta hourglass ($0.9\degree \times 1.4\degree$) and the red rectangle ($1.5\degree \times 0.5\degree$, rotated by $45\degree$). In (b), a graphical representation of the stimulus protocol is depicted comprising two crucial components: first, the background of the stimuli underwent alternating black-and-white transitions following a binary pseudo-random sequence; second, diverse-colored shapes were presented within the stimuli. The stimulus background could dynamically change with each frame of 16.67\,ms (60\,Hz), while the shapes within the stimuli changed every 250\,ms (4\,Hz). A trial took 20~s, within which target shapes (the magenta hour glass) appeared randomly in the sequence with at least 1\,s distance. Participants engaged with the stimuli by counting the number of target shapes on the attended side. In this pilot study, we adopted a paradigm where only the background of the attended stimulus alternated, while the background of the unattended stimulus remained constant. A left-attended trial is shown in (b).}
\end{figure*}

% Stimuli
The stimulus protocol (see Fig.~\ref{fig:stimuli}) was displayed on a 27\,in Corsair Xeneon 27QHD240 OLED screen at a $1920 \times 1080$\,px resolution with a 120\,Hz refresh rate. The participants were seated at a 60\,cm distance in front of the display. A black fixation cross was presented at the center of the screen on a mean luminance gray background. To each of the sides of the fixation cross at a distance of $2.1\degree$, two circles with a $3\degree$ diameter were presented. 
%At the left top of the screen, a black box was placed that turned white at the start of stimulation, triggering a synchronization signal via an opto-sensor that marks the onset of stimulation in the EEG data. The opto-sensor was placed directly on top of the black box such that it was not visible to the participant during the trials. 
% To each of the sides of the fixation cross at a distance of $2.1\degree$\,visual degrees, two circles with a 3~visual degrees diameter were presented. At the left top of the screen, a black box was placed that turned white at the start of stimulation, triggering a synchronization signal via an opto-sensor that marks the onset of stimulation in the EEG data. The opto-sensor was placed directly on top of the black box such that it was not visible to the participant during the trials.

\captionsetup[subfigure]{position=top, labelfont=bf,textfont=normalfont,singlelinecheck=off,justification=raggedright}

% Bit sequences
The circles' background color was luminance modulated with binary pseudo-random noise-codes, such that ones represent a white and zeros a black background. We used 126-bit modulated Gold codes~\cite{gold1967, thielen2015}, which contained only short flashes of 16,67\,ms (bit sub-sequence `010') and long flashes of 33,33\,ms (bit sub-sequence `0110') at a presentation rate of 60\,Hz. From the available modulated Gold codes, we carefully selected one for the left side. For the right circle a 61\,bits phase-shifted version of the left code was used. This was done such that the noise-codes' properties were identical, but still had a near-zero correlation at a maximum delay. 

% Shape sequences
% height and width instead of area
% for rectangle-- exact hieght and width with rotation
Inside the circles ($3\degree$ diameter), five colored shapes were presented with a maximum possible height and width of $1.4\degree$ each. The shapes and their colors are as follows : a green circle ($1.4\degree$ diameter), magenta hourglass ($0.9\degree \times 1.4 \degree$), cyan inverted triangle ($0.9\degree \times 1.4\degree$), red rectangle ($1.5\degree \times 0.5\degree$, rotated by $45\degree$), and the yellow triangle ($0.9\degree \times 1.4\degree$). All shapes had the same brightness and were sequentially presented in random order at a rate of 4\,Hz (see Fig.~\ref{fig:stimuli}). Participants were asked to count the number of times that the magenta hourglass, i.e., the target shape, occurred on the cued side, to facilitate sustaining their attention and to evaluate the behavioral performance of attending to each of the sides. Within a trial, the temporal distance between the presentation of two target shapes was at least 1\,s, the target shape could not be presented on both sides simultaneously, and the number of times the target shape was presented differed for the two sides within a trial. 

% The target to target distance was at least 1\,s, no target could be presented in both circles at the same time, and the number of targets presented at each of the sides was always different within one trial.

% Sequential stimulation 
In this pilot study, we used sequential stimulation in both overt and covert runs to make the first step towards gaze-independent c-VEP BCI. That is, only the circle on the attended (cued) side underwent alternating background changes based on the pseudo-random noise-code, while the unattended side retained a constant black background. Notably, both sides featured distinct shape sequences despite this sequential stimulation protocol. 

\mysubsection{Experiment}
During the experiment, participants completed five runs: four runs required covert attention and one required overt attention, the order of which was randomized across participants. Each run consisted of 20\,trials, 10 for each of the two classes, in random order. At the start of a run, a 5\,s period was used to let the participant prepare for the upcoming trials. At the start of a trial, a 1-second cue was presented to indicate the to-be-attended side using an arrow. 
Subsequently, for a duration of 20\,s, the cued circle flashed according to its bit sequence while the uncued circle remained static, while both circles showed their distinct shape sequences. At the end of a trial, participants were given a maximum of 5\,s to enter the number of target shapes they counted on the attended side using a keyboard, after which they received feedback for a period of 1\,s on the correctness of their response. Finally, before continuing to the next trial, a 1\,s blank inter-trial interval was presented. At the end of a run, the behavioral accuracy of correct responses was shown on the screen. Participants could take self-paced breaks in between runs.

In summary, we gathered 20 trials for each participant in the overt condition, whereas the covert condition involved the recording of 80 trials per participant. In both conditions, the labels (left and right) were balanced.

\mysubsection{Analysis}
% Trial
We used a template-matching classifier to predict the attended side (left or right) given the recorded brain activity. Specifically, we used the `reconvolution' method~\cite{thielen2015}, which assumes that the evoked response to a stimulus sequence can be described by the linear superposition of the responses to the individual flashes in that sequence. The reconvolution approach can substantially reduce the number of parameters while increasing the number of samples to train these parameters, which effectively can limit the required training data~\cite{thielen2021}.

% Event matrix
In reconvolution, the event time-series of the $i$th stimulus sequence are listed in the event matrix $\mathbf{E}_i \in \mathbb{R}^{E \times T}$ for $E$-many events and $T$-many samples. This matrix describes the onset of each of the events in a sequence. In this work, the events were defined as the onset of the stimulation sequence in each trial, and one event for each of the the flash durations (short `010' and long `0110'), for a total of $E=3$ events. 

% Structure matrix
The event time-series are subsequently transformed to a structure matrix that not only describes the onset, but also the modeled length and importantly the overlap of the transient responses for each of the events in the event matrix. Assuming that the transient response length can be limited to $L$ samples without losing relevant data, the structure matrix of the $i$th stimulus sequence is a Toeplitz-like matrix $\mathbf{M}_i \in \mathbb{R}^{M \times T}$ for $M=E*L$ event time points. 

% CCA fit
Let's assume we have a training dataset $\{(\mathbf{X}_1, y_1), (\mathbf{X}_j, y_j) \dots, (\mathbf{X}_J, y_J)\}$ including labeled EEG data for $j \in \{1, ..., J\}$ trials with the single-trial EEG $\mathbf{X} \in \mathbb{R}^{C \times T}$ of $C$-many channels and $T$-many samples and the associated binary label $y \in \{0, 1\}$. With this data, we can learn a spatial filter $\mathbf{w} \in \mathbb{R}^C$ and a temporal response vector $\mathbf{r} \in \mathbb{R}^M$ by maximizing the following correlation $\rho$ as part of a canonical correlation analysis (CCA):
\begin{equation}\label{eq:cca_fit}
    \underset{\mathbf{w}, \mathbf{r}}{\arg\max}~\rho(\mathbf{w}^\top\mathbf{S}, \mathbf{r}^\top\mathbf{D})
\end{equation}
where $\mathbf{S} = [\mathbf{X}_1, \mathbf{X}_j, \dots, \mathbf{X}_J]$ are the concatenated single trials and $\mathbf{D} = [\mathbf{M}_{y_1}, \mathbf{M}_{y_j}, \dots, \mathbf{M}_{y_J}]$ are the concatenated accompanying structure matrices. 

% CCA predict
Having learned the spatial filter and temporal response vector, we can now predict the label of a new trial $\hat{y}$ by maximizing the following Pearson's correlation $\rho$:
\begin{equation}\label{eq:cca_predict}
    \hat{y} = \underset{i}{\arg\max}~\rho(\mathbf{w}^\top\mathbf{X}, \mathbf{r}^\top\mathbf{M}_i)
\end{equation}
Here, $\mathbf{w}^\top\mathbf{X}$ is the spatially filtered data and $\mathbf{r}^\top\mathbf{M}_i$ is the predicted response template for the $i$th stimulus sequence.

% Evaluation
To evaluate the performance of the reconvolution CCA on the overt and covert data, we used a chronological 4-fold cross-validation within each condition. The classification accuracy was averaged across folds. Note, the c-VEP stimulation was only applied on the attended side, while the unattended side remained a black background color. In the decoding analysis, we simulated as if the unattended side had been flashing with the noise-code other than the one presented on the attended side.

% Code
Code for the reconvolution CCA approach is available at: \url{https://github.com/thijor/pyntbci}.

%----------------------------------------------------------------------------------------
% Results
%----------------------------------------------------------------------------------------

\mysection{results}

\begin{figure*}[h]
    \centering
    \includegraphics[width=\textwidth]{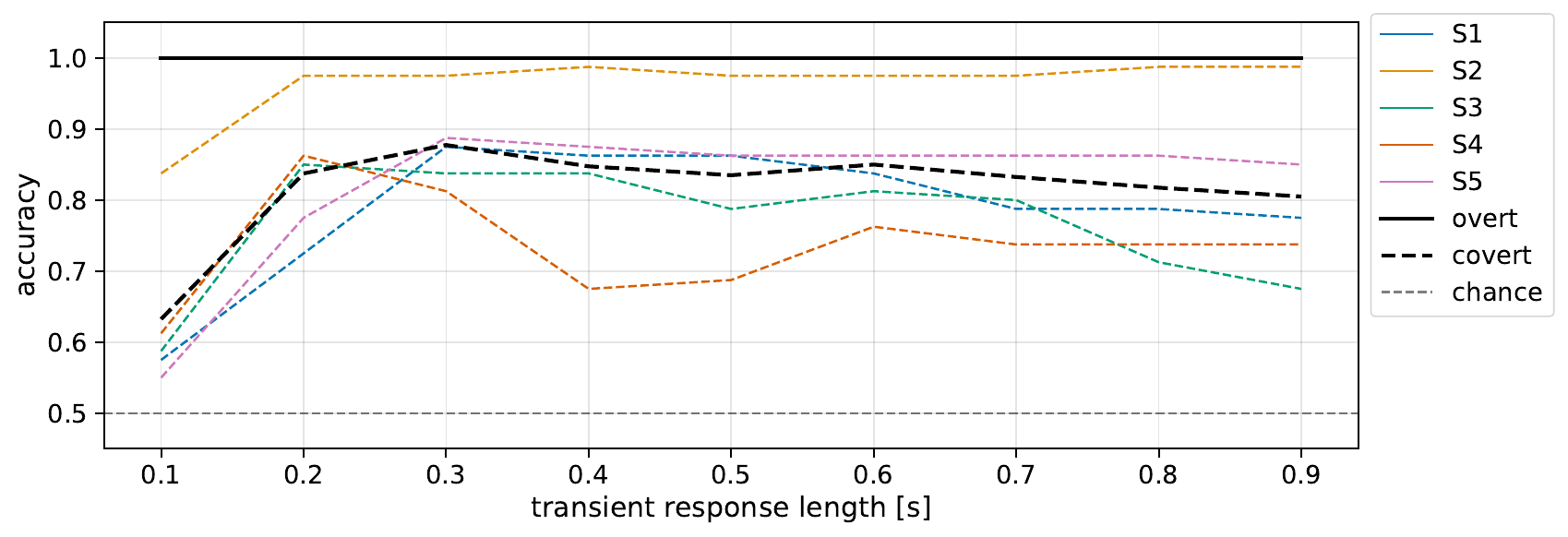}
  \caption{\label{fig:accuracy}\textbf{Classification accuracy across modeled transient response lengths.} Depicted are the participant-specific classification accuracies for both overt (solid lines) and covert (dashed lines) conditions across transient response lengths ranging from 0.1\,s to 0.9\,s. The grand average over participants is shown in black. Please note, that for the overt condition, the classification accuracy was $100$\,\% for all transient response lengths and all participants. The dashed gray line indicates theoretical chance level ($50$\,\%).}
\end{figure*}

As this study presents the initial step to decode c-VEP from peripheral stimulation, aiming towards covert spatial attention, it is imperative to study how classification accuracy is influenced by the modeled transient response length. Given the potential for distinct transient responses between conditions, we assessed the mean accuracy across transient response lengths spanning from 0.1 to 0.9\,s, for all participants (S1-S5) and both conditions (see to Fig.~\ref{fig:accuracy}).

In the covert condition, mean accuracy fluctuated from $55$\,\% to $99$\,\% across participants, whereas in the overt condition, mean accuracy remained consistently at $100$\,\% for all participants across all transient response lengths.

In the covert condition, participants S3 and S4 achieved a peak accuracy of $85$\,\% and $86$\,\% respectively, observed at a transient response length of 200\,ms. Participants S1 and S5 reached a highest accuracy of $88$\,\% and $89$\,\%, respectively, at a transient response length of 300\,ms. Participant S2 demonstrated a peak accuracy of $99$\,\% at 400\,ms. Notably, the mean accuracy across participants in the covert condition was highest at a transient response length of 300\,ms. Hence, for subsequent analysis, we use a transient response length of 300\,ms.

 Tab.~\ref{tab:accuracy} shows the classification accuracy for a transient response length of 300\,ms. The scores obtained in the covert condition for S1-5 were $88\,\%, 98\,\%, 84\,\%, 81$\,\% and $89$\,\%, respectively, leading to an average of $88$\,\%. The overt condition performed better for all participants ($100$\,\%). All individual scores in Tab.~\ref{tab:accuracy} are significantly higher ($p < .001$) than chance level ($50$\,\%) as verified by a permutation test using $1000$ permutations.

\begin{table}[h]
    \centering
        \caption{\label{tab:accuracy}\textbf{Mean classification accuracy.} The table shows the classification accuracy using a transient response length of 300\,ms, for each participant and the grand average, for both overt and covert conditions. All classification results for both conditions and all participants individually were significantly higher than chance ($50$\,\%) as verified by a permutation test with 1000 permutations ($p < .001$).}
    \begin{tabular}{l|rrrrr|r}\toprule
        & \textbf{S1} & \textbf{S2} & \textbf{S3} & \textbf{S4} & \textbf{S5} & \textbf{Avg} \\\midrule
        \textbf{Overt}   & 1.00    & 1.00    & 1.00    & 1.00    & 1.00     & 1.00 \\
        \textbf{Covert}  & 0.88    & 0.98    & 0.84    & 0.81    & 0.89    & 0.88 \\\bottomrule
    \end{tabular}
\end{table}
% use width of 11.69/2 

\begin{figure*}
    \centering
    \begin{subfloat}[]{
        \includegraphics[width=\columnwidth]{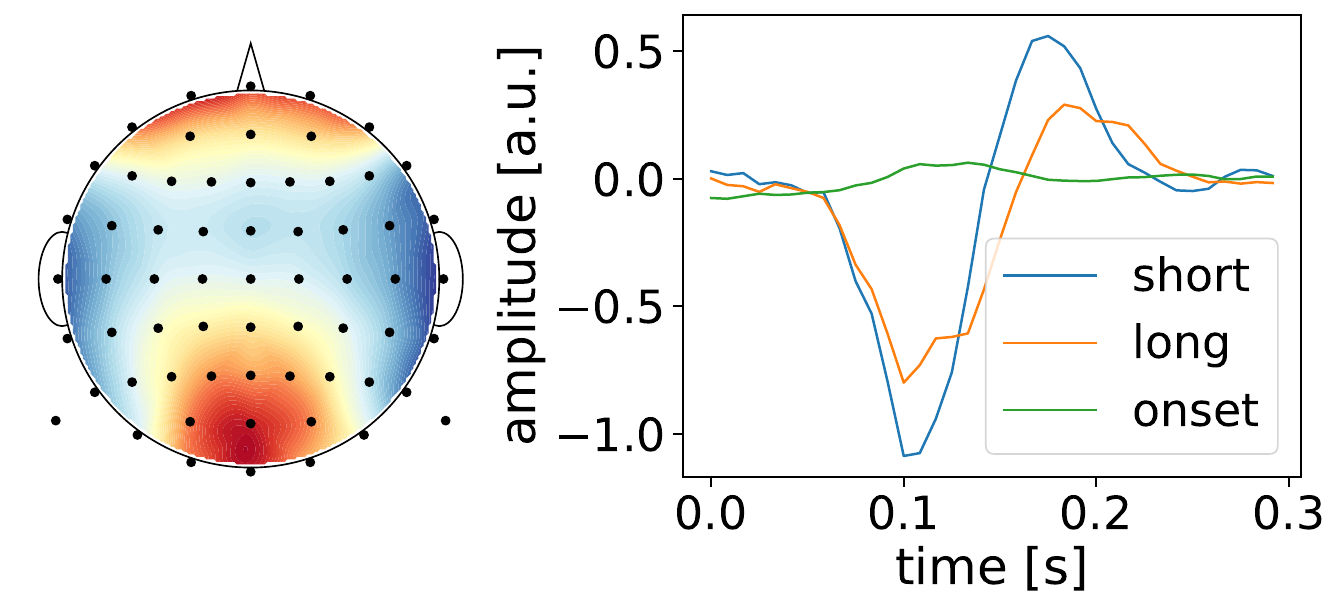}
    }
    \end{subfloat}
    \begin{subfloat}[]{
        \includegraphics[width=\columnwidth]{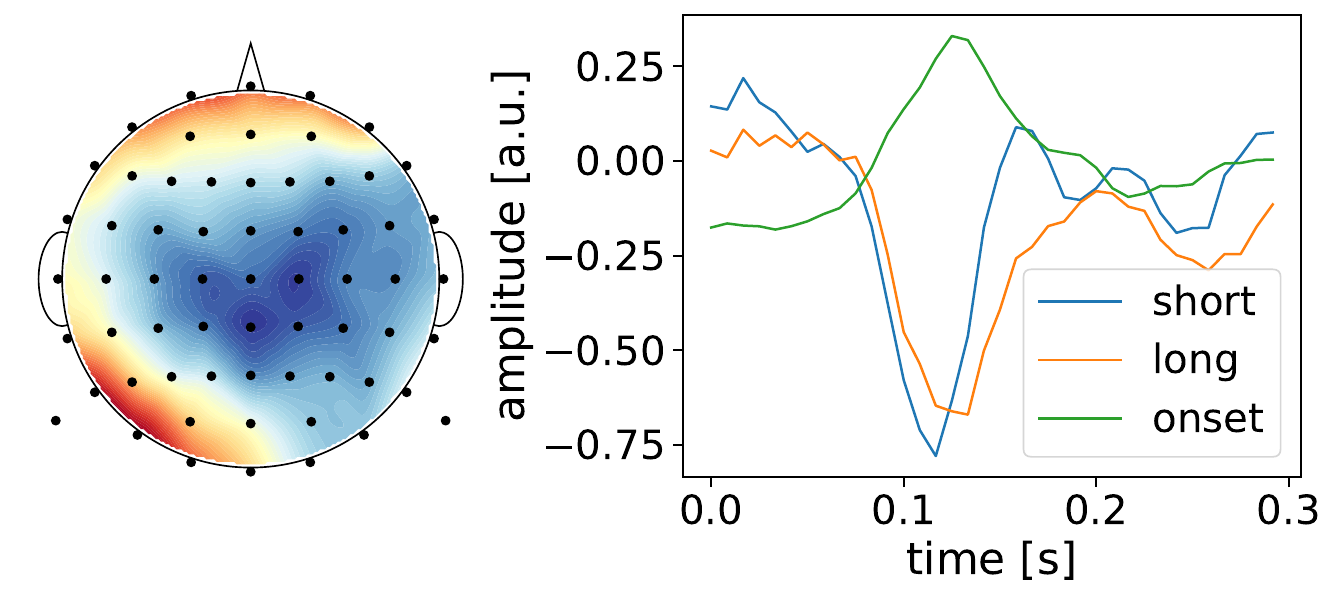}
    }
    \end{subfloat}
     \caption{\label{fig:responses}\textbf{Spatial activity pattern and transient responses of participant S4.} (a) and (b) show the spatial activity pattern and transient responses of S4 for the overt and covert conditions, respectively. For all participants, the spatial activity for the overt condition was more focally distributed as compared to the more lateralized distribution seen for the covert condition. The spatial pattern $\mathbf{a} \in \mathbb{R}^C$ was estimated as $\mathbf{a}=\mathbf{w}^\top{\bm\Sigma}$, where ${\bm\Sigma} \in \mathbb{R}^{C \times C}$ is the spatial covariance matrix.}
\end{figure*}

To investigate the differences in characteristics of the spatial activity patterns and transient responses, we computed these at a transient response length of 300\,ms for both conditions. Fig.~\ref{fig:responses} shows an example of the spatial pattern and transient responses for S4. Across participants, we observed that the spatial activity pattern for the overt condition was more focally distributed, whereas it was more lateralized for the covert condition. 

%----------------------------------------------------------------------------------------
% Discussion
%----------------------------------------------------------------------------------------

\mysection{discussion}

% Aims and design
Our pilot study provides fundamental insights into the plausibility of a c-VEP-based stimulation paradigm for decoding covert spatial attention, thereby potentially eliminating the need for the ability to make eye movements to control a c-VEP BCI. We implemented a two-class paradigm, requiring participants to attend on a stimulus either to the left or to the right of their fixation point. The stimuli background flashed following pseudo-random noise-codes, while their foreground simultaneously presented a random sequence of five distinct shapes with an infrequent target shape. Participants were tasked with counting the occurrences of the target shape amidst the shape sequence (see Fig.~\ref{fig:stimuli}). In this pilot study, we used sequential stimulation to assess the feasibility of covert c-VEP, before moving to the more complex parallel stimulation requiring covert spatial attention.

% Summary of results
In our experiment, participants engaged with the stimuli through either overt means, involving eye movements to foveate on the target, or covertly, relying on spatial attention to focus on a target. In the overt condition, we reached a decoding performance of $100$\,\% for all participants. In the covert condition, we achieved an average accuracy of $88$\,\%. To the best of the authors' knowledge, this marks the first evaluation of a c-VEP BCI using covert attention, although here we still rely on sequential stimulation. Our study highlights the feasibility of such a design for developing gaze-independent BCIs that can be used by people with ALS.

% Comparison to literature -- compare to Egan; we are higher but mention number of participants and sequential stimulation
In the overt condition, all participants achieved 100\,\% accuracy, likely caused by the large data availability, low number of classes, and sequential stimulation. Specifically, this study used 5.3\,min of data for training and 20\,s for testing, while 1\,min training and 1-2\,s testing would suffice~\cite{thielen2021}. 
In the covert condition, we employed 16\,min of data for training, achieving a decoding accuracy of $88$\,\%. This result underscores the lower SNR in the covert condition compared to the overt scenario. Nevertheless, although using sequential stimulation, the attained performance surpasses the $62$\,\% accuracy reported in a similar SSVEP study that used parallel stimulation~\cite{egan2017}, offering evidence for the potential performance of gaze-independent c-VEP.

% Limitations
It is essential to approach the results of our study on gaze-independent c-VEP BCI with caution and consider two important limitations. Firstly, this preliminary study involved a small cohort of five highly motivated participants. Secondly, the c-VEP protocol employed sequential stimulation, where only the stimulus on the attended side alternated its background based on the pseudo-random noise-code. In practical online usage of the BCI, simultaneous stimulation on both sides is necessary. While our study offers valuable fundamental insights into the feasibility of gaze-independent c-VEP BCI, it is imperative to acknowledge these limitations. Further research, including a larger sample size and parallel stimulation, is crucial to fully unveil the potential of this approach.

% Paradigms other than VEP and their performances   
Additionally, it is important to acknowledge that stimulation paradigms outside the visual domain have been explored as well for developing independent BCIs.
For instance, Schreuder and colleagues developed the P300-based auditory multi-class spatial ERP (AMUSE) interface reaching a classification accuracy of about $85$\,\% ($N=6$ classes)~\cite{Schreuder2011}. 
Similarly, Brouwer and van Erp designed a P300-based BCI using vibro-tactile feedback around the waist with an accuracy of $58$\,\% ($N=6$ classes) and $73$\,\% ($N=2$ classes)~\cite{Brouwer2010}. Moreover, Van der Waal and colleagues~\cite{vanderWaal2012} used tactile stimulation on the finger tips reaching a classification accuracy of $82$\,\% ($N=6$ classes). These results may also highlight the potential to explore the pseudo-random stimulation protocol in the auditory and tactile domain.

% Future work
Our studies' design enables the use of two additional features in the analysis pipeline, possibly further improving the accuracy. 
Firstly, the stimulus protocol used in the study was designed such that the infrequent occurrence of the target events within the shape sequence could potentially evoke a P300 response. Hence, the P300 response could be used alongside the c-VEP to decode the attended side, similar to P300 response that was used alongside the SSVEP by Egan and colleagues~\cite{egan2017}. 
Secondly, the alpha-band modulations are expected to be lateralized with respect to the attended side~\cite{worden2000}. Specifically, covertly attending to a stimulus on one side suppresses visual alpha-activity in the contra-lateral (task-positive) hemisphere, while it increases alpha in the ipsi-lateral (task-negative) hemisphere~\cite{jensen2010}. Hence, visual alpha oscillations can also be used as an additional feature, again similar to the alpha response used alongside the SSVEP in earlier work~\cite{egan2017}. 
Thirdly, aligning with the anticipated lateralization in the alpha-band, we also anticipate lateralization in the c-VEP itself during the covert condition. In our current application of reconvolution CCA, a single spatial filter was employed to decode the attended side. This method can be extended by incorporating distinct spatial filters for each side, a concept referred to as an `ensemble' decoder~\cite{gembler2020b}.
Finally, in the present study, we employed only two stimuli positioned on either side of the fixation point, using luminance modulation with two 126-bit Gold codes. Given the limited number of classes, there is potential to explore shorter codes, which could lead to faster decoding. Furthermore, alternative codes, such as the m-sequence or Golay sequence, may be considered, as they have shown promise in enhancing classification accuracy~\cite{thielen2023a}.

% \begin{itemize}
%     \item Show overt and covert performance, with emphasis on higher than chance level decoding in covert. First time shown for c-VEP.
%     \item Limitation: sequential presentation, should be parallel.
%     \item Future: incorporate lateralization in the rCCA method.
%     \item Future: incorporate the P300 and alpha as in Egan work in a hybrid framework.
% \end{itemize}

%----------------------------------------------------------------------------------------
% Conclusion
%----------------------------------------------------------------------------------------

\mysection{conclusion}

Our study shows the feasibility and high performance of a novel covert BCI design based on c-VEP. Our design eliminates the dependence on gaze, which is an essential feature if BCIs are to be used by people that have no voluntary control over their eye movements, such as people living with late stage ALS. Further, the design of the study makes it possible to use additional measures of brain activity to improve classification performance, which is a potential fruitful avenue for future work to improve the efficacy of the gaze-independent c-VEP BCI. Overall, our results suggest the potential for a high-speed BCI that does not rely on any overt behavior.

%----------------------------------------------------------------------------------------
% Acknowledgements
%----------------------------------------------------------------------------------------

\mysection{Acknowledgements}

This work was part of the project `Obtaining fast brain-computer interfacing without eye movements for communication and control' with project number OCENW.XS23.1.127 of the research programme `Open Competitie ENW XS' which is financed by the Dutch Research Council (NWO).

%----------------------------------------------------------------------------------------
% References
%----------------------------------------------------------------------------------------

\mysection{references}
\printbibliography[heading=none]

@article{blankertz2006,
  title={The {Berlin} Brain-Computer Interface presents the novel mental typewriter {Hex-o-Spell}},
  author={Blankertz, Benjamin and Dornhege, Guido and Krauledat, Matthias and Schr{\"o}der, Michael and Williamson, John and Murray-Smith, Roderick and M{\"u}ller, Klaus-Robert},
  year={2006}
}

@article{Brouwer2010,
  title = {A tactile {P300} brain-computer interface},
  ISSN = {1662-453X},
  url = {http://dx.doi.org/10.3389/fnins.2010.00019},
  DOI = {10.3389/fnins.2010.00019},
  journal = {Frontiers in Neuroscience},
  publisher = {Frontiers Media SA},
  author = {Brouwer},
  year = {2010}
}

@inbook{Chen2016,
  title = {An Online Gaze-Independent {BCI} System Used Dummy Face with Eyes Only Region as Stimulus},
  ISBN = {9783319399553},
  ISSN = {1611-3349},
  url = {http://dx.doi.org/10.1007/978-3-319-39955-3_3},
  DOI = {10.1007/978-3-319-39955-3_3},
  booktitle = {Foundations of Augmented Cognition: Neuroergonomics and Operational Neuroscience},
  publisher = {Springer International Publishing},
  author = {Chen,  Long and Allison,  Brendan Z. and Zhang,  Yu and Wang,  Xingyu and Jin,  Jing},
  year = {2016},
  pages = {26–34}
}

@article{egan2017,
  title = {A gaze independent hybrid-{BCI} based on visual spatial attention},
  volume = {14},
  ISSN = {1741-2552},
  url = {http://dx.doi.org/10.1088/1741-2552/aa6bb2},
  DOI = {10.1088/1741-2552/aa6bb2},
  number = {4},
  journal = {Journal of Neural Engineering},
  publisher = {IOP Publishing},
  author = {Egan,  John M and Loughnane,  Gerard M and Fletcher,  Helen and Meade,  Emma and Lalor,  Edmund C},
  year = {2017},
  month = may,
  pages = {046006}
}

@article{gembler2020b,
	title={Asynchronous {c-VEP} communication tools—efficiency comparison of low-target, multi-target and dictionary-assisted {BCI} spellers},
	author={Gembler, Felix W and Benda, Mihaly and Rezeika, Aya and Stawicki, Piotr R and Volosyak, Ivan},
	journal={Scientific Reports},
	volume={10},
	number={1},
	pages={1--13},
	year={2020},
	publisher={Nature Publishing Group},
	doi = {10.1038/s41598-020-74143-4},
	url = {https://doi.org/10.1038/s41598-020-74143-4}
}

@article{gold1967,
  title = {Optimal binary sequences for spread spectrum multiplexing},
  volume = {13},
  ISSN = {1557-9654},
  url = {http://dx.doi.org/10.1109/TIT.1967.1054048},
  DOI = {10.1109/tit.1967.1054048},
  number = {4},
  journal = {IEEE Transactions on Information Theory},
  publisher = {Institute of Electrical and Electronics Engineers (IEEE)},
  author = {Gold,  R.},
  year = {1967},
  month = oct,
  pages = {619–621}
}

@article{jensen2010,
  title={Shaping functional architecture by oscillatory alpha activity: gating by inhibition},
  author={Jensen, Ole and Mazaheri, Ali},
  journal={Frontiers in Human Neuroscience},
  volume={4},
  pages={186},
  year={2010},
  publisher={Frontiers Research Foundation}
}

@article{kelly2005,
  title = {Visual spatial attention tracking using high-density {SSVEP} data for independent brain-computer communication},
  volume = {13},
  ISSN = {1558-0210},
  url = {http://dx.doi.org/10.1109/TNSRE.2005.847369},
  DOI = {10.1109/tnsre.2005.847369},
  number = {2},
  journal = {IEEE Transactions on Neural Systems and Rehabilitation Engineering},
  publisher = {Institute of Electrical and Electronics Engineers (IEEE)},
  author = {Kelly,  S.P. and Lalor,  E.C. and Reilly,  R.B. and Foxe,  J.J.},
  year = {2005},
  month = jun,
  pages = {172–178}
}

@article{martnezcagigal2021,
  title = {Brain–computer interfaces based on code-modulated visual evoked potentials ({c-VEP}): a literature review},
  volume = {18},
  ISSN = {1741-2552},
  url = {http://dx.doi.org/10.1088/1741-2552/ac38cf},
  DOI = {10.1088/1741-2552/ac38cf},
  number = {6},
  journal = {Journal of Neural Engineering},
  publisher = {IOP Publishing},
  author = {Martínez-Cagigal,  Víctor and Thielen,  Jordy and Santamaría-Vázquez,  Eduardo and Pérez-Velasco,  Sergio and Desain,  Peter and Hornero,  Roberto},
  year = {2021},
  month = nov,
  pages = {061002}
}

@article{thielen2015,
  title = {Broad-Band Visually Evoked Potentials: Re(con)volution in Brain-Computer Interfacing},
  volume = {10},
  ISSN = {1932-6203},
  url = {http://dx.doi.org/10.1371/journal.pone.0133797},
  DOI = {10.1371/journal.pone.0133797},
  number = {7},
  journal = {PLOS ONE},
  publisher = {Public Library of Science (PLoS)},
  author = {Thielen,  Jordy and van den Broek,  Philip and Farquhar,  Jason and Desain,  Peter},
  editor = {Marinazzo,  Daniele},
  year = {2015},
  month = jul,
  pages = {e0133797}
}

@article{thielen2021,
	title={From full calibration to zero training for a code-modulated visual evoked potentials for brain--computer interface},
	author={Thielen, Jordy and Marsman, Pieter and Farquhar, Jason and Desain, Peter},
	journal={Journal of Neural Engineering},
	volume={18},
	number={5},
	pages={056007},
	year={2021},
	publisher={IOP Publishing},
	doi = {10.1088/1741-2552/abecef},
	url = {https://doi.org/10.1088/1741-2552/abecef}
}

@InProceedings{thielen2023a,
	author={Thielen, Jordy},
	editor={Rojas, Ignacio and Joya, Gonzalo and Catala, Andreu},
	title={Effects of Stimulus Sequences on Brain-Computer Interfaces Using Code-Modulated Visual Evoked Potentials: An Offline Simulation},
	booktitle={Advances in Computational Intelligence},
	year={2023},
	publisher={Springer Nature Switzerland},
	address={Cham},
	pages={555--568},
	isbn={978-3-031-43078-7},
	doi = {10.1007/978-3-031-43078-7_45},
	url = {https://doi.org/10.1007/978-3-031-43078-7_45}
}

@article{treder2010,
  title={({C})overt attention and visual speller design in an {ERP}-based brain-computer interface},
  author={Treder, Matthias S and Blankertz, Benjamin},
  journal={Behavioral and Brain Functions},
  volume={6},
  pages={1--13},
  year={2010},
  publisher={Springer}
}

@article{treder2011a,
  title = {Gaze-independent brain–computer interfaces based on covert attention and feature attention},
  volume = {8},
  ISSN = {1741-2552},
  url = {http://dx.doi.org/10.1088/1741-2560/8/6/066003},
  DOI = {10.1088/1741-2560/8/6/066003},
  number = {6},
  journal = {Journal of Neural Engineering},
  publisher = {IOP Publishing},
  author = {Treder,  M S and Schmidt,  N M and Blankertz,  B},
  year = {2011},
  month = oct,
  pages = {066003}
}

@article{treder2011b,
  title={Brain-computer interfacing using modulations of alpha activity induced by covert shifts of attention},
  author={Treder, Matthias S and Bahramisharif, Ali and Schmidt, Nico M and Van Gerven, Marcel AJ and Blankertz, Benjamin},
  journal={Journal of Neuroengineering and Rehabilitation},
  volume={8},
  number={1},
  pages={1--10},
  year={2011},
  publisher={BioMed Central}
}

@article{Schreuder2011,
  title = {Listen, You are Writing! {Speeding} up Online Spelling with a Dynamic Auditory {BCI}},
  volume = {5},
  ISSN = {1662-4548},
  url = {http://dx.doi.org/10.3389/fnins.2011.00112},
  DOI = {10.3389/fnins.2011.00112},
  journal = {Frontiers in Neuroscience},
  publisher = {Frontiers Media SA},
  author = {Schreuder, Martijn and Rost, Thomas and Tangermann, Michael},
  year = {2011}
}

@article{shi2024,
  title={Estimating and approaching the maximum information rate of noninvasive visual brain-computer interface},
  author={Shi, Nanlin and Miao, Yining and Huang, Changxing and Li, Xiang and Song, Yonghao and Chen, Xiaogang and Wang, Yijun and Gao, Xiaorong},
  journal={NeuroImage},
  pages={120548},
  year={2024},
  publisher={Elsevier}
}

@article{vanderWaal2012,
  title = {Introducing the tactile speller: an {ERP}-based brain–computer interface for communication},
  volume = {9},
  ISSN = {1741-2552},
  url = {http://dx.doi.org/10.1088/1741-2560/9/4/045002},
  DOI = {10.1088/1741-2560/9/4/045002},
  number = {4},
  journal = {Journal of Neural Engineering},
  publisher = {IOP Publishing},
  author = {van der Waal,  Marjolein and Severens,  Marianne and Geuze,  Jeroen and Desain,  Peter},
  year = {2012},
  month = jul,
  pages = {045002}
}

@article{verbaarschot2021,
	title={A visual brain-computer interface as communication aid for patients with amyotrophic lateral sclerosis},
	author={Verbaarschot, Ceci and Tump, Dani{\"e}lle and Lutu, Andreea and Borhanazad, Marzieh and Thielen, Jordy and van den Broek, Philip and Farquhar, Jason and Weikamp, Janneke and Raaphorst, Joost and Groothuis, Jan T and others},
	journal={Clinical Neurophysiology},
	volume={132},
	number={10},
	pages={2404--2415},
	year={2021},
	publisher={Elsevier},
	doi = {10.1016/j.clinph.2021.07.012},
	url = {https://doi.org/10.1016/j.clinph.2021.07.012}
}

@article{waytowich2015,
	title={Spatial decoupling of targets and flashing stimuli for visual brain--computer interfaces},
	author={Waytowich, Nicholas R and Krusienski, Dean J},
	journal={Journal of Neural Engineering},
	volume={12},
	number={3},
	pages={036006},
	year={2015},
	publisher={IOP Publishing},
	doi = {10.1088/1741-2560/12/3/036006},
	url = {https://doi.org/10.1088/1741-2560/12/3/036006}
}

@article{wolpaw2002,
  title = {Brain–computer interfaces for communication and control},
  volume = {113},
  ISSN = {1388-2457},
  url = {http://dx.doi.org/10.1016/S1388-2457(02)00057-3},
  DOI = {10.1016/s1388-2457(02)00057-3},
  number = {6},
  journal = {Clinical Neurophysiology},
  publisher = {Elsevier BV},
  author = {Wolpaw,  Jonathan R and Birbaumer,  Niels and McFarland,  Dennis J and Pfurtscheller,  Gert and Vaughan,  Theresa M},
  year = {2002},
  month = jun,
  pages = {767–791}
}

@article{worden2000,
  title = {Anticipatory Biasing of Visuospatial Attention Indexed by Retinotopically Specific $\alpha$-Bank Electroencephalography Increases over Occipital Cortex},
  volume = {20},
  ISSN = {1529-2401},
  url = {http://dx.doi.org/10.1523/JNEUROSCI.20-06-j0002.2000},
  DOI = {10.1523/jneurosci.20-06-j0002.2000},
  number = {6},
  journal = {The Journal of Neuroscience},
  publisher = {Society for Neuroscience},
  author = {Worden,  Michael S. and Foxe,  John J. and Wang,  Norman and Simpson,  Gregory V.},
  year = {2000},
  month = mar,
  pages = {RC63–RC63}
}

\end{document}